# Statistical Interpretation of 'Femto-Molar' Detection

Jonghyun Go & Muhammad A. Alam

*School of Electrical and Computer Engineering, Purdue University, West Lafayette, IN 47907, USA*

**Over the last decade, many experiments have demonstrated that nanobiosensors based on Nanotubes and Nanowires are significantly more sensitive compared to their planar counterparts. Yet, there has been persistent gap between reports of analyte detection at ~femto-Molar concentration and theory suggesting the impossibility of sub-pM detection at the corresponding incubation time. This divide has persisted despite the sophistication of the theoretical models. In this paper, we calculate the statistics of diffusion-limited arrival-time distribution by a Monte Carlo method to suggest a statistical resolution of the enduring puzzle: The incubation time in the theory is the *mean* incubation time, while experiments suggest device stability limited the *minimum* incubation time. The difference in incubation times – both described by characteristic power-laws – provides an intuitive explanation of different detection limits anticipated by theory and experiments. These power laws broaden the scope of problems amenable to the 'first-passage process' used to quantify the stochastic biological processes.**

In recent years, the development of nanobiosensors based on nanowires and nanotubes have allowed rapid and remarkable progress in ultrasensitive detection of chemicals and biomolecules. While different groups report widely different detection numbers, publications from several leading groups [1-4] suggest that detection limit may approach femto-Molar concentration – establishing a level of sensitivity for nano-biosensors matched by few other analytical techniques. Given the additional prospect of label-free detection, top-down fabrication and ease of integration, it is easy to understand the excitement for this technology.

Subsequent theoretical analysis have been only partially successful in interpreting the experimental results and this absence of a experimentally validated theoretical framework have made it difficult to compare results from different laboratories and have frustrated rapid optimization of the technology. In general, theoretical models based on diffusion-limited capture of for analyte molecules do suggest that 'geometry of diffusion' makes cylindrical nanowires superior to planar FET sensors in their ability to respond to low analyte density. Specifically it has been shown in Ref. [5] that the average incubation time $t_{avg}$ for a given analyte density $\rho_0$ follows the scaling relationship $\rho_0 t_{avg}^{M_D} \sim k_D$ where $M_D$ and $k_D$ are sensor-geometry dependent constant, and the time exponent $M_D$ is equal to 1 for the one-dimensional nanowire sensor [4]. Given the specific dimension of the sensors (defines $k_D$) used in experimental demonstration of 'femto-Molar detection' and given reported incubation time of few minutes, the scaling law suggests a theoretical lower limit of detection of $\rho_s \sim 1$ pM – with no obvious explanation of the gap between theory and experiments. Additional consideration involving analyte transport in fluid flow [6], improves the diffusion-limited detection limit by possibly a factor 10. The use of electrokinetic approaches to improve local concentration has been suggested, but given that the strong screening at typical mM salt concentration and appreciating that 1 fM translates to 1 molecule in $10^6$ $\mu m^{-3}$, diffusion-limited detection limit unlikely to be modified appreciably by these techniques [7]. Moreover, fM detection has been observed with variety of sensor configurations and flow conditions, thereby suggesting the possibility of simpler and more robust explanation of the observed phenomena.

In this paper, we offer a statistical interpretation to resolve this puzzle (see Fig. 1). The theoretical models of biosensors [5,6,9] consider the response of an asymptotically large system (e.g., a NW of infinite length) and as such, the predicted theoretical response time is relevant for practical nanobiosensors only in the sense of an ensemble average, i.e., time when ~50% for sensors in a large sensor array registers the presence of the analytical molecule. In practice, the finite size of NW sensors dictates that some element of the ensemble (sensor) would respond before the others, i.e. the response time will be statistically distributed. Given the number of sensors in an array is finite and their lifetime in harsh fluid environment limited [8] (See Sec. B of Appendix for a detailed discussion of the stability-limited response time), the reported response times could actually be the minimum response time of the system – representing the tail of a broad arrival time distribution. Fig. 1 shows that for a given incubation time, the requirement that only one or few sensors respond compared to that of requirement that 50% of the sensors respond could lead to significantly different detection limits for a given technology. Fig. 1a shows a generalized picture of a nanobiosensor in which a number of analyte particles diffuse inside a domain before being absorbed by a box (or sensor). If we assume that the box generates a response only after it captures a given numbers of molecule (Fig. 1b), then it is easy to see that the incubation time for signal generation will be stochastically distributed with a specific probability distribution (Fig. 1c) and that this distribution will depend on the density of the analyte molecules. If we plot the minimum and average detection times ($t_{min}$ and $t_{avg}$) as a function of density (Fig. 1d), we find that they might be considerably different – with a possible resolution of the puzzle of the minimum detection limit for a given sensor technology (Fig. 1e).

Consideration of the aforementioned hypothesis requires calculation of arrival time distribution for different analyte densities in the regime of diffusion-limited transport. For this purpose, let us consider a cylindrical nanowire sensor surrounded by a static analyte solution. The specific receptors for the target molecules are immobilized on the surface of the sensor. In the Reaction-Diffusion(R-D) model [10], the conjugation dynamics of target molecules to their receptors is described by

$$\frac{dN_s(t)}{dt} = k_f(N_0 - N_s(t))\rho_s(t) - k_r N_s(t) \tag{1}$$

where $N_s$ is the density of conjugated receptors, $N_0$ is the density of receptors on the sensor surface, $k_f$ and $k_r$ are the capture and dissociation constants. The concentration of molecules at the sensor surface at a given time $t$, $\rho_s(t)$, is determined by R-D model as well as by the diffusion of target molecules set by the concentration gradient at the sensor surface which is given by the diffusion equation

$$\frac{\partial \rho(\vec{r},t)}{\partial t} = D\nabla^2 \rho(\vec{r},t) \tag{2}$$

where $D$ is the diffusion coefficient of target molecules in the solution. The classical solution of Eqs. (1) and (2) provides 'ensembled-averaged' response time of biosensors

In the following discussion, we assume the sensor is described by a perfectly absorbing boundary condition ($k_f \to \infty, k_r = 0$). This assumption implies that transport is diffusion-limited (Damkohler number > 1) and that the biosensor response time at a particular concentration (to be defined below) is shorter than the reaction limited saturation time (i.e., $t < t_S \left[ = 1/(k_r + k_f \rho) \right]$). The validity of both these assumptions are discussed in Sections B and C of Appendix.

To calculate distribution of response/registration times – not simply their ensemble-average -- one must determine sample-specific response of biosensor at a given analyte density by solving Eq. (1) and (2) stochastically by Monte Carlo (MC) method. The direct Monte Carlo solution is computationally intractable – explaining why no such calculation has ever been reported in the literature despite its broad interest and obvious relevance for large class of stochastic biological problems (e.g., [12-14]). Instead of using a direct Monte Carlo method, we use the following novel variant of the MC technique – the so-called to "Table-based MC (TMC) approach" [15,16] – to analyze the problem. The basic idea of TMC is to use the MC method to numerically calculate and tabulate the capture time distributions for particles injected at various starting position, i.e. to numerically precalculate and store the Green's function [12] from any random starting point to the sensor surface. For example, the (*i,j*)-th element of the table, $G_{i,j} \equiv G(r_i, t_j)$ describes the probability that a particle injected at location $r_i$ at time *t=0* is captured by the sensor at time $t_j = j\Delta t$. For the calculation proper, a sample "S" is first created by specifying the initial positions of the particles ( $r_k^S; k=1,\cdots,M$ ) consistent with a specific density of analyte, $\rho_M$. For each particle from $r_k^S$, its capture time by the sensor is stochastically chosen to be consistent with the pre-calculated arrival time distribution from that point, $G_{i,()} \equiv G(r_i,...)$. The process is repeated for all *M* analyte molecules of the sample "S" to obtain a sorted list of arrival times, $t_{s,m}$; $m=1...M$. If *k* is the number of particles required for an observable sensor response, then $t_{s,m=k}$ is the initial response time for this sensor. The process is repeated for large number of samples (N~1000s) to establish an *k*-th arrival-time distribution at a particular density of analytes $\{t_{s=1...N;k}\}$. The ensemble average of $\{t_{s=1...N;k}\}$ coincides with the continuum solution of (1) and (2), as expected. We have also independently verified that this approach correctly reproduces the analytical results for many-particle capture dynamics for simplified cases of 1D diffusion (see section E in Appendix).

For an illustrative example (see Figs. 2-3), we consider a NW of radius 50nm [1], the length 1μm, diffusion coefficient $D \sim 10^{-6}$ cm$^2$/s (For a summary of various geometrical and physical parameters, see Appendix, table 1), simulation lattice size $\Delta x$ = 20nm and simulation time increment $\Delta t$ is 1μs. Fig. 2 shows the normalized distributions of the *k*-th arrival times (*k* = 1,3,5) for an ensemble of 2,000 NW-based biosensors (*N* = 2000). Specifically, the red-line indicates PDF of arrival times for those sensors sensitive enough to register the presence of analyte by the capture of a single molecule (*k* = 1). Similarly, for sensors that require *at least* three analyte molecules to be captured before registering an output detection signal (*k* = 3), the blue line describes the PDF of registration times by various sensors in the array. Several features of the distribution are obvious: First, for all PDFs, the average arrival time (~50% of the sensors indicating the presence of analyte) is significantly larger compared to the corresponding minimum arrival time when – for example – 5-10% of the sensors indicates the presence of the analyte, promising a resolution of the 'theory-experiment' gap discussed in the introduction. Second, the ratio of $t_{k,\min} / t_{k,\text{avg}}$ reduces with $k$ – in other words, more sensitive the sensor ($k$ ~ smaller), larger is the gap between $t_{k,\min}$ and $t_{k,\text{avg}}$. Finally, the minimum of the $k$-th arrival times exhibit different scaling laws from the average arrival time, because the statistical probability of $k$ molecules being placed close to the sensors is different from average response that is dictated by distribution of all samples.

Let us now examine the minimum and average of the arrival times among 2,000 biosensor requiring at least 5 molecules to register an output signal (i.e., PDF of $t_{N=2000;k=5}$ with $k = 5$) and measure the variations of minimum response times with respect to various analyte concentrations. Remarkably, analogous to the scaling law for average response time $\rho_0 t_{\text{avg}}^{M_D} \sim k_D$ or equivalently $t_{\text{avg}} \propto \rho^{M_D^{-1}}$, Fig. 3 suggests the following simple scaling law of the minimum response time for detecting $k$ molecules:

$$t_{k,\min} = t_0 \left(\frac{\rho_0}{\rho}\right)^{\alpha_k} \quad (k=1,2,\cdots) \tag{3}$$

where $\alpha_k$ (see Fig. 4a) is the power exponent of the minimum time to detect target molecules.

Fig. 3a shows that for biosensors with single-molecule sensitivity ($k = 1$), the minimum arrival time increases much more slowly compared the average response. This difference of response time with $\rho_0$ is easy to understand: As the concentration of analyte molecules is reduced, the average distance of molecules from the sensor increases rapidly and so does the average response time. At sub-nM concentration, the minimums of the *k*-th arrival times ($k \geq 1$) increases with the same-power exponent as the average response time (~1, see Fig. 4a), since the probability that multiple molecules are populated at the very next to the sensor surface decrease as rapidly as the average distance of molecules away from the sensor surface. This implies that the ratio of the minimum to the average of the *k*-th arrival times at a low density :

$$t_{k,\min} = c(k) \, t_{k,\text{avg}} \qquad (4)$$

where $t_{k,\text{avg}}$ indicates the average response time to detect *k* target molecules derived from the scaling law. In Fig. 4b we find that *c*(*k*) increases rapidly at the low arrival order but increases slowly at the high arrival order, and eventually it could approach to 1 when the arrival order goes to infinity, as expected. The value of $c(k) \sim 10^{-2}$–$10^{-3}$ for low values of *k* indicates that the theoretically calculated average arrival time reported in the literature could differ from the experimentally relevant minimum arrival time by a 2-3 orders of magnitude. Since $\alpha_k \sim 1/M_D \sim 1$, the difference in arrival time directly translates into the difference in minimum detection limits. Specifically, in Fig. 3b, we find that the minimum detection limit corresponding to typical incubation time of ~100 seconds decreases by more than 3 orders of magnitude when one compares the minimum and average response time curves for *k* = 1. This provides a simple resolution of the gap between previous theoretical results (~pM) and experimental demonstrations (~fM). Obviously, the gap reduces rapidly at higher density – explaining why such an issue has not been dominant for older classical sensors. In general, Eq. (4) may be used to simply estimate the minimum response time for molecular detection first using the scaling law of the average response time, $\rho_0 t_{\text{avg}}^{M_D} \sim k_D$, and then multiplying the ratio *c*(*k*) to the average response time.

Given the difference between minimum and average incubation times and corresponding difference in detection limits, it is important to provide technology-specific context of such limits. For example, for a sensors network widely dispersed in a battlefield to signal the presence of a single bio-agent (i.e., nerve gas), the specification of minimum response time is relevant because the registration of the molecules with first few sensors is sufficient to trigger system-wide response. Average response time is irrelevant for this application because one need not wait till 50% of the sensors to have responded before corrective actions can be taken. On the other hand, for sensors arrays involving in proteomic and geomonic applications, all the sensors must complete bindings before the experiment is terminated. In this case, the relevant incubation time is much larger than even average response time at that concentration. Therefore, the minimum response time – when only a few sensors of the array completing the conjugation process – is an irrelevant indicator of promise/utility of the biosensor technology for such technological applications. Under this circumstance, the only viable option of detecting ~fM analyte might involve using a larger ensemble of nanowire sensors per pixel to rapid average the statistical signal [4].

To conclude, we have developed a comprehensive model describing the statistical distributions of response time for an ensemble of cylindrical NW/NT biosensors. Our numerical studies suggest that the minimum and average response are both characterized by respective power laws, and at low analyte concentration, their magnitude could differ by more than 2-3 orders of magnitude depending on the minimum number of molecules ($k$ values) needed to activate the sensor. This implies that it is possible to detect target chemicals or biomolecules within time less than we estimate from the classical diffusion-limited scaling laws [5,6,9] of average response time or mean first passage time, and our new scaling law provides a framework for interpretation of experimental detection at femto-Molar concentration, thereby resolving an enduring puzzle in the biosensor literature. Additionally, our model is closely related to the problems regarding the first passage process [11,12] or narrow-escape time [13,14] in absorbing boundaries and it is easy to extend and generalize our approach to solve the $k$-th passage process among $N$ particles for those biologically relevant transit times with broad applications.

Acknowledgements   We thank Pradeep R. Nair for detailed discussions. This work was supported by funds from the Network for Computational Nanotechnology (NCN) and National Institute of Health (NIH).

Author Information   Correspondence and requests for materials should be addressed to J. Go or M. A. Alam (go@purdue.edu, alam@purdue.edu).

**Figure 1** (**a**) The generalized model of nanobiosensors: absorbing box inside a domain captures diffusing particles and (**b**) generates a detection signal once a minimum of $k$ particles has been captured. Robust positive detection is indicated when $N_c$ such sensors (out of a total of $N_{sample}$) captures at least $k$ particle each. The shadowed region indicates a noise of the signal. (**c**) The population distribution of the response times for different molecular densities: $\rho_1 > \rho_2$. One can correspondingly define their minimum response time ($t_{min}$) as being the $N_c$th-smallest one among those $N_{sample}$ response times. $p_c$ indicates the pdf for t = $t_{min}$ whose corresponding cdf is $N_c / N_{sample}$. (**d**) The statistical nature of diffusion process makes the minimum response time (dotted line) significantly different from the 'ensemble-average' time (solid line). (**e**) For a given incubation time $T_{inc}$, the detection limit based on minimum arrival time ($\rho_{DL}$)' is significantly lower than that based on the classical ensembled average detection limit, $\rho_{DL}$.

**Figure 2** Population distributions of the first ($k = 1$), third ($k = 3$), and fifth ($k = 5$) arrival time for a nanowire biosensor at a molecular concentration $\rho$ = 53 nM. $k$ = 1 implies sensors capable of single molecule detection, while $k > 1$ implies less sensitive sensors that requires multiple analyte capture for positive detection above the noise floor.

**Figure 3** (**a**) The scaling relationship of both the minimum and average of the $k$-th arrival time with respect to the molecular concentration for a nanowire-based biosensor. The distributions of arrival times are obtained by collecting the $k$-th arrival times from 2,000 samples of biosensor, i.e. $N_{sample}$=2000. Note that

$N_C$=20, corresponding to 1% tail of the PDF. (See Appendix D for responses for other definitions of the tail of the PDF). (**b**) From an extrapolated curve of the first arrival time ($k$ = 1, solid green curve), the minimum detection limit corresponding to a given incubation time decreases by about 3 order of magnitude. Dotted curves indicate the average response times for $k$ = 1,…5.

**Figure 4** (**a**) Power exponent $α_k$ for the minimum of the $k$-th arrival time. (**b**) The ratio $c(k)$ with respect to the arrival order $k$ at various molecular densities demonstrates the universality of the scaling function based on Fig. 3 and 4.

Figure 1.

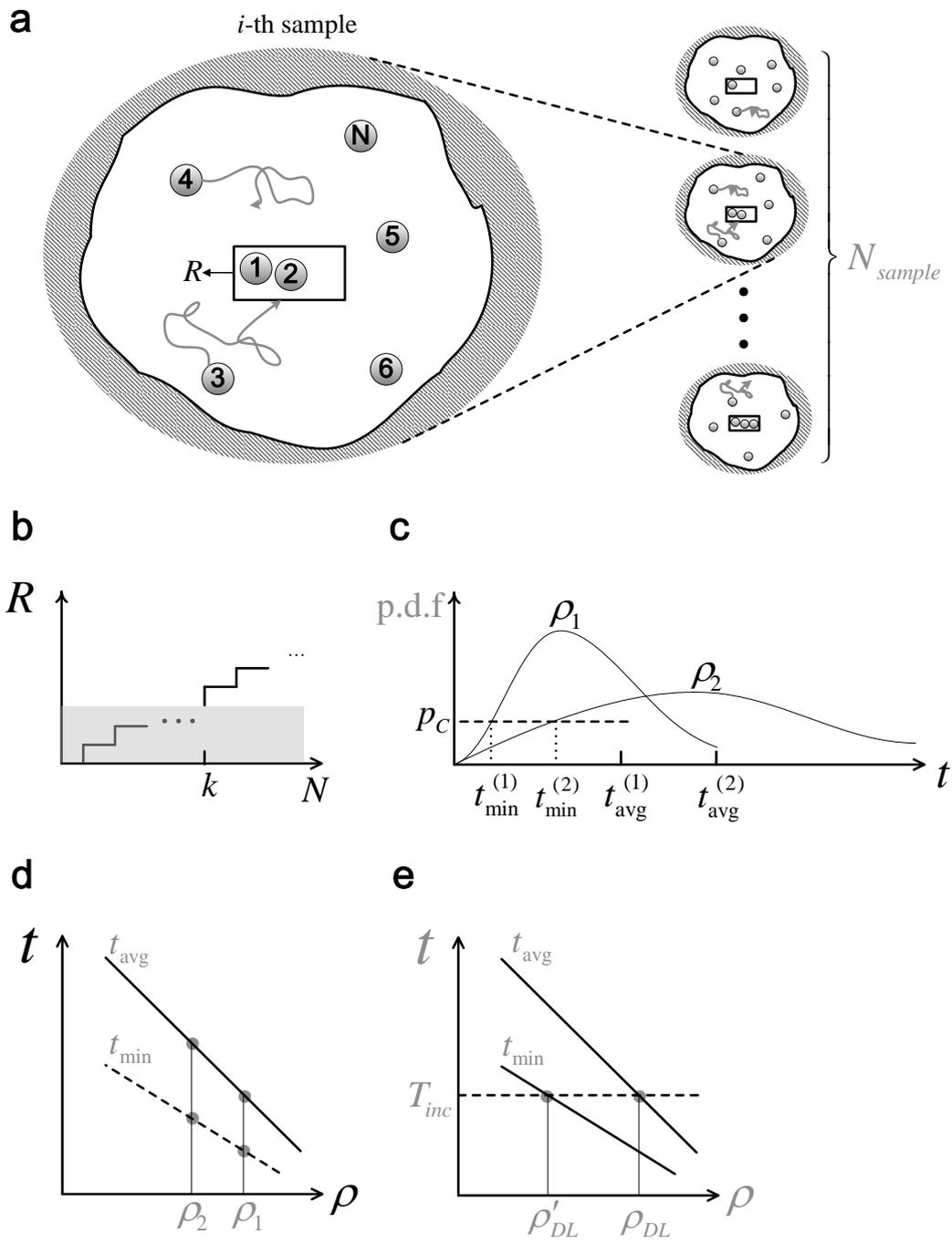

Figure 2.

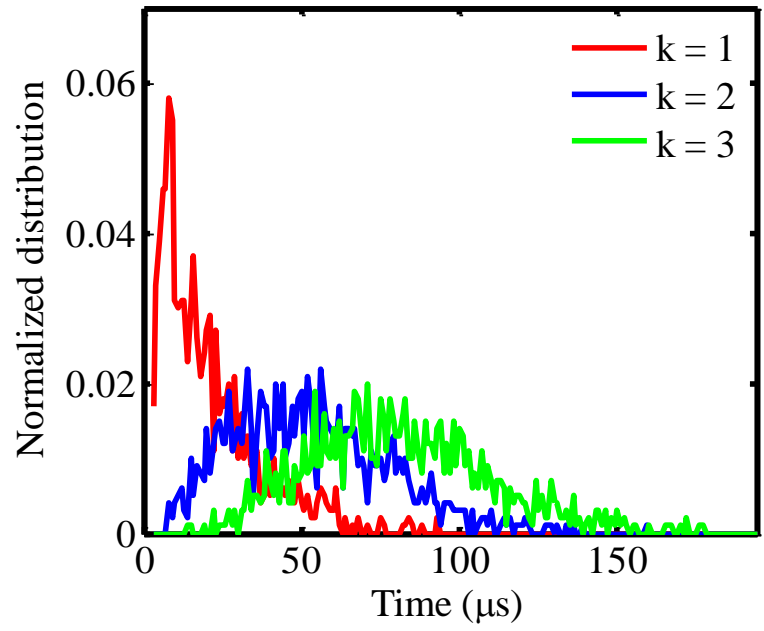

Figure 3.

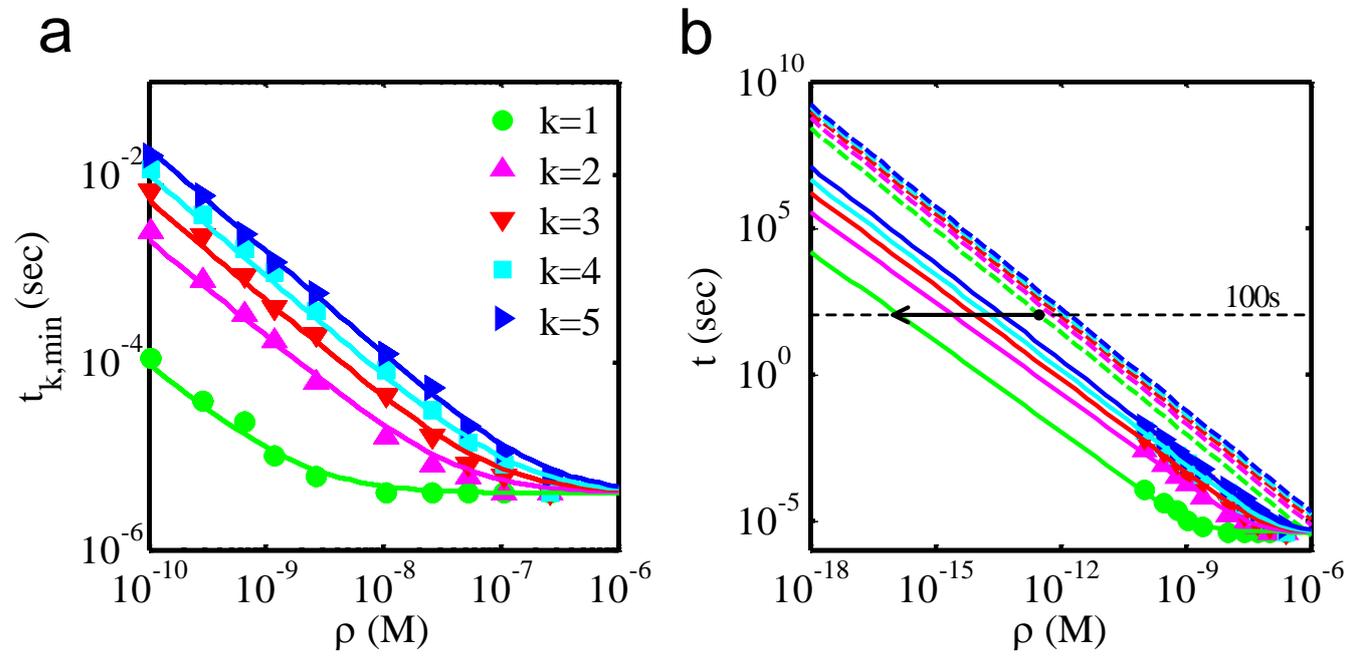

Figure 4.

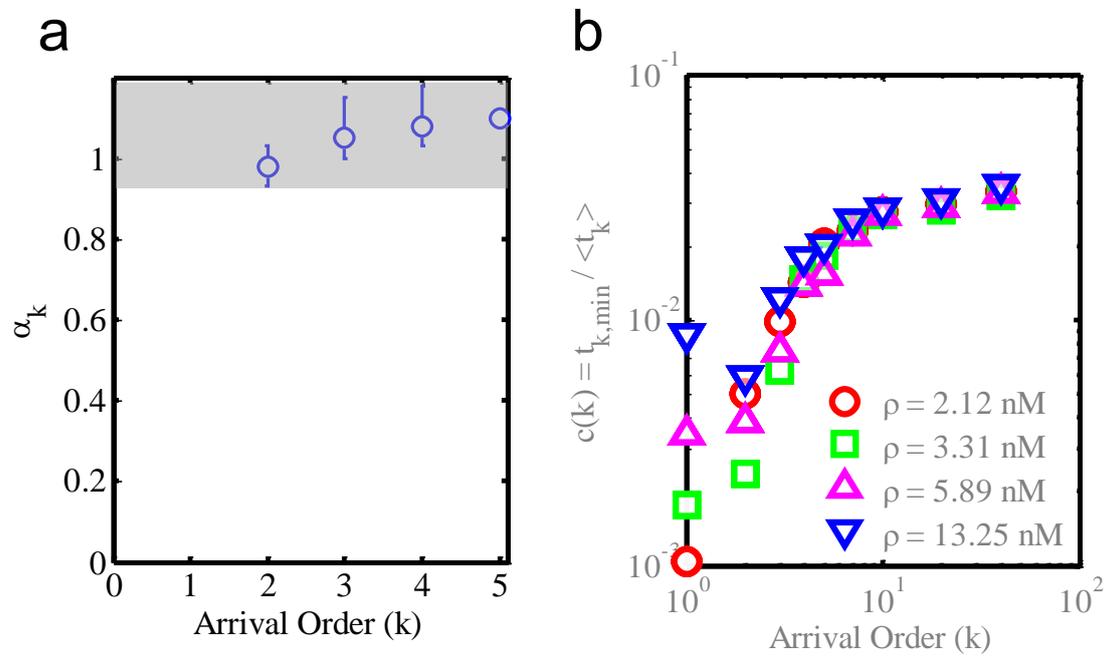

# Appendix

# Statistical Interpretation of 'Femto-Molar' Detection

J. Go, P. R. Nair, and M. A. Alam

*School of Electrical and Computer Engineering, Purdue University, West Lafayette, IN 47907, USA*

### A. Diffusion vs. Reaction Limited Transport

In this paper, we have presumed that diffusion or mass-transport dominates the biosensor response, even in case of finite volumetric flow rate Q ~ 10 μL/min. To justify this assumption, we should show that the Damkohler numbers (defined as the ratio of the diffusion flux to reaction flux) of the relevant experiments are greater than 1. Based on the reported geometrical dimensions and physical parameters for various experiments in the literature, Table 1 shows that indeed Da > 1 for all cases, thereby justifying an important assumption in the paper.

*Calculation Procedure*. Squires et al. (Ref. [S8]) and Sheehan et al. (Ref. [S9]) suggest two different approaches for calculating the diffusion flux ($J_D$) for a given value of volumetric flow rate Q. Both provide reasonable results for sub-100 nm NWs, although Ref. [S9] compares better with numerical solution at higher flow rates and NWs with larger diameters (see Appendix Figure 1 below). The fluxes ($J_{D,1}$, $J_{D,2}$) and Damkohler numbers ($Da_1$, $Da_2$) -- based on Ref. [S8] and Ref. [S9] respectively -- are calculated as follows:

$$J_{D,1} = D\rho_0 W_S F \quad \text{and} \quad Da_1 = k_f b_m L / DF.$$

$$J_{D,2} = DW_S \rho_0 \frac{2\pi}{4.885 - \ln(P_S)} \left(1 - \frac{0.09266 P_S}{4.885 - \ln(P_S)}\right) \quad \text{where} \quad P_S = \frac{6QL^2}{DW_C H^2}$$

$$Da_2 = k_f \rho_0 b_m L W_S / J_{D,2}$$

|  | Physical Parameters | | | Calculated Da Numbers | |
|---|---|---|---|---|---|
|  | $L$ (nm) [a] | $D$ (cm$^2$/s) | $k_f$ (M$^{-1}$s$^{-1}$) | $Da_1$ [b] | $Da_2$ [c] |
| Ref. [S1] | ~20 | $1.6 \times 10^{-8}$ [S10] | $3 \times 10^{8}$ [S10] | 622.5 | 624.0 |
| Ref. [S2] | ~40 | $8.5 \times 10^{-7}$ [S11] | $3 \times 10^{7}$ [d] | 5.028 | 5.025 |
| Ref. [S3] | ~50 | $8.2 \times 10^{-7}$ [e] | $3 \times 10^{7}$ [d] | 4.0593 | 4.0570 |
| Ref. [S4] | ~50 | $8.2 \times 10^{-7}$ [e] | $3 \times 10^{7}$ [d] | 4.7307 | 4.7274 |
| Ref. [S5] | ~100 | $4 \times 10^{-7}$ [S13] | $3 \times 10^{7}$ [d] | 15.199 | 15.195 |

Table 1. Calculation of Damkohler Numbers (Da) (a) $L$ indicates the diameter of a given nanowire. We have taken the average of the values reported in the corresponding references. (b) Damkohler numbers derived from the formulas suggested in Ref. [S8]. (c) Damkohler numbers derived from the formulas proposed in Ref. [S9]. (d) The estimation of $k_f$ is based on Ref. [S10]. (e) The diffusion coefficient is calculated based on assumption of 12 base-pairs DNA (Ref. [S12]). Most DNA experiments involve longer segments that would correspondingly reduce the diffusion coefficient and increase the Da numbers.

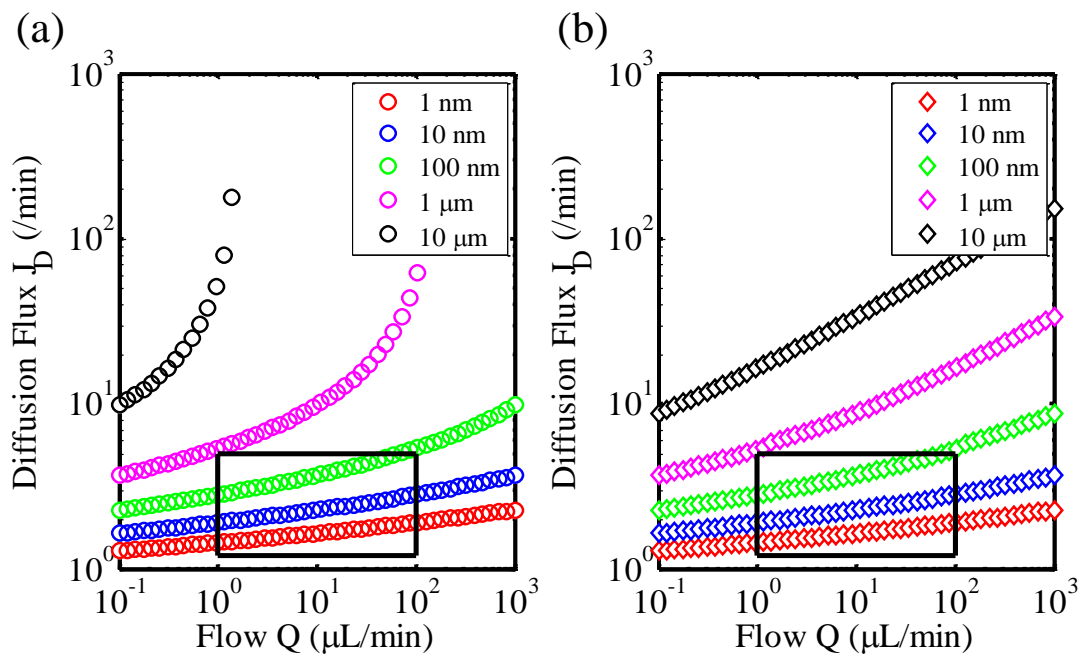

Appendix Figure 1. The diffusion flux of molecules with respect to the volumetric flow rate Q for several different nanowire widths is computed using two approaches. Both of (a) Squires' approach (Ref. [S8]) and (b) Sheehan's approach (Ref. [S9]) shows quite similar values of $J_D$ in our interested regime of Q around 10 µL/min. Here we used the same device dimensions and physical parameters as Sheehan suggested.

## B. Considerations of 'Device Stability Time' in interpreting Theoretical Response Limits of Biosensors

In Appendix Figure 2 below, we summarize the experimental results (symbols) reported from wide variety of publications in the literature. The response times were either obtained directly from those reported in the paper, or obtained indirectly by fitting the experimental time-response characteristics with theoretical time-response. In contrast to Fig. 1 in the manuscript, we show only one set of theoretical mean (dashed line) and minimum (solid line) response curves for visual clarity.

In order to interpret the experiments in the light of the theory proposed in this paper, we note that the *mean* response time (dashed line) is meaningful only if the devices remain electrically stable up to the theoretically predicted mean response time. The salt environment, quality of thin oxides, and various fabrication details, however, result in a broad range of stability times for nanobiosensors. For example, Ref. [S5] suggests stability time of ~30 sec (Fig. 2, dashed black line), while Ref. [S3] shows that the devices remain stable for many hours (Fig. 5). In general, the typical stability time of sensors is limited to 5-30 min (R. Bashir, UIUC, personal communication).

It is clear from the data from Refs. [S3], [S4], and [S7] that whenever the device stability time exceeds the theoretical response time and whenever average data from several devices are reported (e.g., Ref. [S3] reports average of 30-60 samples), the data follows the theoretical *mean* diffusion-limited response time reasonably well (filled symbols).

In contrast, if the stability time is less than the average response time (Ref. [S5], open symbols), the theory presented in this paper would suggest that measured response would be limited to the statistical tail of the distribution. This theory therefore provides a simple and consistent interpretation of the observed time responses for various reports in the literature.

Other data from the literature [S1,S2,S6] do not provide sufficient information about stability time or sample size (filled light-blue symbols) making it difficult to compare them to the prediction of the stochastic arrival time theory presented in the paper. However, they data do indicate the variability of response time for same analyte concentration, indicating perhaps that statistical origin of these response curves. And the finite response times of several minutes may be dictated by the general stability time (~5-30 min) of such devices.

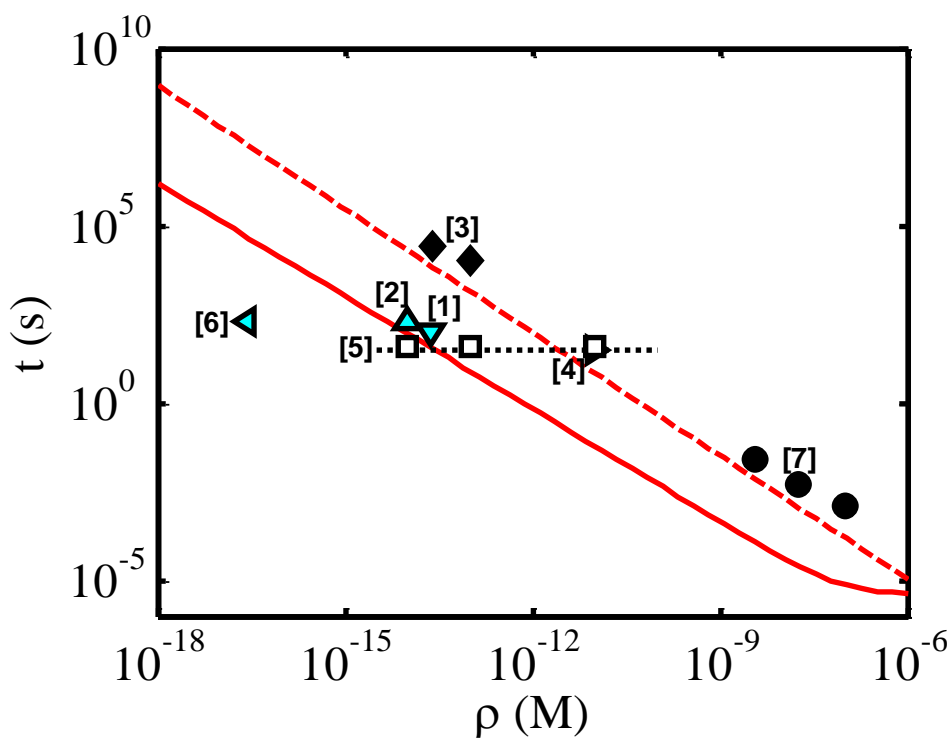

Appendix Figure 2. The detection times with their corresponding analyte concentration reported from several papers (See References). The red solid and dashed lines represent the minimum and average response time to detect 3 molecules (k = 3), respectively.

Ref. [S1]: p.53, Figure 3B, curve 3 (10fM), detection time ~ 200 s.
Ref. [S2]: p.1296, Figure 2a, injection number 4 (0.9 pg/ml PSA = 24 fM), detection time ~100 s.
Ref. [S3]: p.3295, Figure 5, curves 2 (25 fM, tau ~ 173 min) and 3 (100 fM, tau ~ 445 min).
Ref. [S4]: p.1260, Figure 5, stage III (10 pM). detection time ~ 30 s.
Ref. [S5]: p.521, Figure 3d (10 fM, 100 fM, and 10 pM).
    Please look at the fifth line of the left column, p. 521. '*device instability observed ~30 s*'.
Ref. [S6]: p.103901-2, FIG. 2a. (1 fg/ml PSA ~ 27 aM), detection time ~ 200 s.
Ref. [S7]: p.799, Figure 4B. (The three datapoints are scaled downward by a constant factor only to highlight the diffusion-limited response of the system.)

## C. Considerations of Reaction Limited Saturation Times in interpreting Theoretical Response Limits of Biosensors

In this section, we examine the saturation time of molecular conjugation in the reaction-limited regime and compare it to our diffusion-limited model. Appendix Fig. 3a below shows two kinds of sensor responses for different sets of reaction/dissociation constants: (i) sensors with perfect and irreversible binding ($k_f = \infty$, $k_r = 0$; Eq. 1 of the manuscript), and (ii) biosensors with finite $k_f$ (=$4\times10^7$ $M^{-1}s^{-1}$; Table 1 above) and $k_r$ (~ $4\times10^{-7}$ $s^{-1}$ [S14]) values. We also assume that the conjugation of ~10 molecules on a single nanowire surface generates a noticeable change in conductance [S16] (consistent with our manuscript where k = 1~5), thus the corresponding density of conjugated molecules for a nanowire with 1 μm length and 50 μm radius is ~ 100 $um^{-2}$ or $10^{10}$ $cm^{-2}$. It is clear that so long the saturation time $t_S = 1/(k_r + k_f \rho)$ is greater than the response times (intercept points indicated by circles), both approaches provide identical response times. As such, the simpler diffusion-limited transport assumed for the numerical simulation provides quantitatively correct estimates of density-dependent response times.

For clarity, let us illustrate the same point in a slightly different manner. Note that Fig. 3 of the manuscript was calculated with the assumption of perfect and irreversible binding ($k_f = \infty$, $k_r = 0$). We reproduce for reference two such curves from Fig. 3 for mean and minimum response times in the Appendix Fig. 3b below (red lines). The corresponding saturation times $t_s$ for finite $k_f = 4\times10^7$ $M^{-1}s^{-1}$ and $k_r$ ~$4\times10^{-7}$ $s^{-1}$ [S14] calculated from Appendix Figure 3a are now replotted as solid blue line in Appendix Figure 3b. Since the criteria $t < t_S$ is satisfied down to ~150 aM, our use of the idealized response to interpret fM response is indeed acceptable. Note that although higher value of $k_r$ (~ $10^{-3}$ $s^{-1}$) are sometimes used by various groups for illustrative purposes [S8], the values of $k_r$ ~$4\times10^{-6}$ - $4\times10^{-7}$ are consistent with well calibrated melting point measurements and calculations [S14, S15].

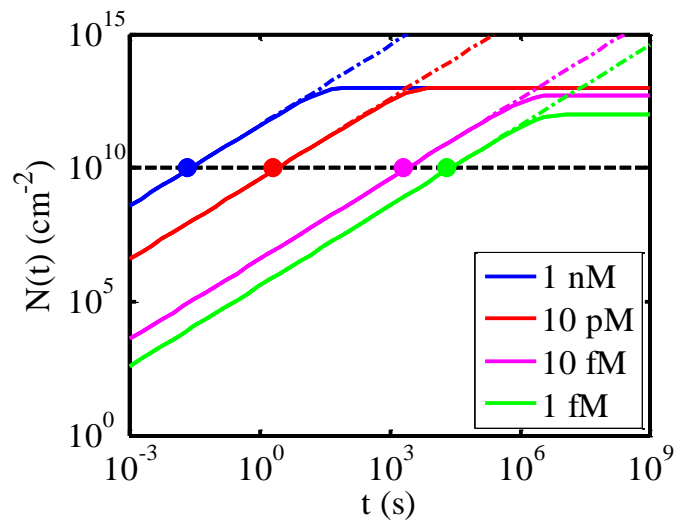

Appendix Figure 3a: Sensor response for a constant supply system, with two different reaction / dissociation rates. Dashed line are average response for a system with $k_f=\infty$, $k_r=0$, and $N_0\sim\infty$ and the solid lines indicate response for finite $k_f$ (= $4\times10^7$ $M^{-1}s^{-1}$) and $k_r$ (= $4\times10^{-7}$ $s^{-1}$) and $N_0$ (=$10^{13}$

cm$^{-2}$). The transition of the responses from linear to saturated region is marked by the saturation time, $t_s(\rho)$.

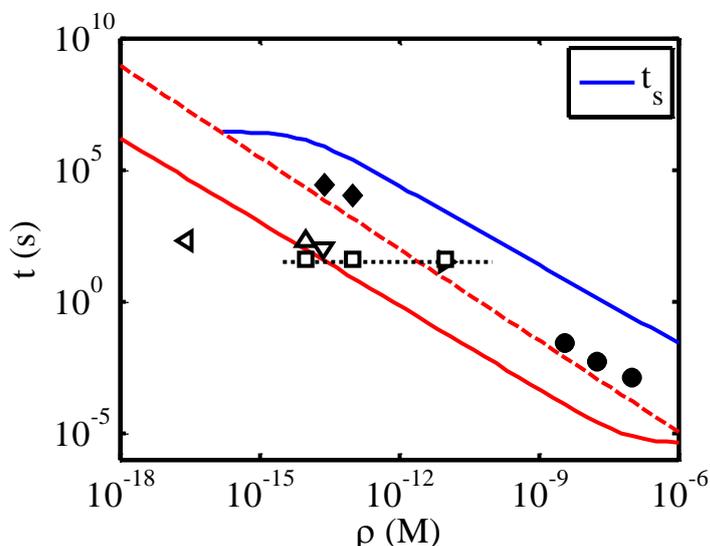

Appendix Figure 3b: The dashed and solid red curves represent the average and minimum response time shown in Fig. 3 of the manuscript. The solid blue curve indicates the reaction-limited staturation time with the association constant ($k_f$) and dissociation constant ($k_r$) coming from Table 1 and Ref. [S14], respectively. The idealized response is accurate so long $t_s \geq t$.

**D. Criterion for Minimum Response Based on the Tail of the Arrival Time Distribution**

In Fig. 3 of the manuscript, we have used the 1% tail of the arrival time distribution to indicate the minimum response time. This 1% criterion is arbitrary and was chosen purely for illustrative purposes. In practice, the limit could be 5-10% depending on technology, device stability time and analyte density. Changing the limit from 1% (few devices having stable characteristics out of several hundred) to 10% (few devices having stable characteristics out of tens of sensors) changes the precise numerical values of response time, but as Appendix Figure 4 below shows it does not affect the breadth of arrival time distribution (ratio of the average time to the minimum response time) significantly.

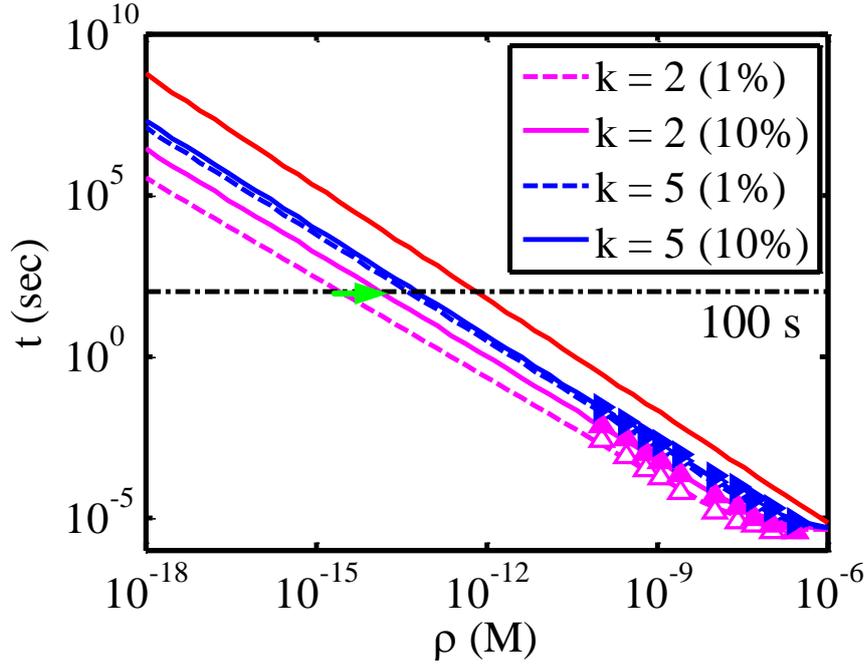

Appendix Figure 4. The shifts of minimum response times with various definitions of experimentally relevant tail of the distribution: Dashed lines and solid lines represent the 1% and 10% tails of the whole response times with two different arrival orders (k = 2, 5), respectively (The red solid line represents the average response time). The amount of shift in detection limit for k=2 curves are indicated as the green arrow, while there is little shift for k=5 curves.

### E. $k^{th}$ passage process of N diffusing particles: Theory and Modeling

Here we derive the analytic expressions for $k^{th}$ passage process among $N$ particles in a one-dimensional diffusion system. Then we compare them to our numerical results obtained by the table-based Monte-Carlo method to validate the approach we suggested in this paper.

I. First-passage time of diffusing $N$ particles in a 1-D semi-infinite system

We start from the following equation for the one-dimensional diffusion equation :

$$\frac{\partial c(x,t)}{\partial t} = D\frac{\partial^2 c(x,t)}{\partial x^2} . \tag{1}$$

Let us assume that a diffusing particle starts at $x_0 > 0$ in a one-dimensional semi-infinite system under the absorbing boundary condition that the *normalized* concentration $c(x,t)$ at the origin is zero. The initial and boundary condition of this system is given by

$$c(0,t) = 0 \tag{2a}$$

$$c(x,0) = \delta(x-x_0) \tag{2b}$$

Solving the equation (1) subject to the conditions in equation (2) gives us [S17]

$$c(x,t) = \frac{1}{\sqrt{4\pi Dt}}[e^{-(x-x_0)^2/4Dt} - e^{-(x+x_0)^2/4Dt}] \tag{3}$$

First of all, we define the survival probability $S_0(t)$ as the probability that a diffusing particle starting at $x_0$ has not hit the absorbing boundary by time $t$. From its definition $S_0(t)$ can be found by integrating the normalized concentration $c(x,t)$ over all $x>0$, then we obtain

$$\begin{aligned} S_0(t) &= \int_0^\infty c(x,t)dx \\ &= \frac{1}{\sqrt{4\pi Dt}}\left[\int_0^\infty e^{-(x-x_0)^2/4Dt}dx - \int_0^\infty e^{-(x+x_0)^2/4Dt}dx\right] \\ &= \frac{1}{\sqrt{4\pi Dt}}\sqrt{4Dt}\left[\int_{-\frac{x_0}{\sqrt{4Dt}}}^\infty e^{-\xi^2}d\xi - \int_{\frac{x_0}{\sqrt{4Dt}}}^\infty e^{-\xi^2}d\xi\right] \\ &= \frac{1}{\sqrt{\pi}}\int_{-\frac{x_0}{\sqrt{4Dt}}}^{\frac{x_0}{\sqrt{4Dt}}} e^{-\xi^2}d\xi = \frac{2}{\sqrt{\pi}}\int_0^{\frac{x_0}{\sqrt{4Dt}}} e^{-\xi^2}d\xi = \mathrm{erf}\left(\frac{x_0}{\sqrt{4Dt}}\right) \end{aligned} \tag{4}$$

Next, we introduce another variable $f_0(t)$: a density function of the probability that a diffusing particle starting at $x_0$ visits the origin for the first time at time $t$. By integrating $f_0(t)$ with respect to time one can calculate the probability that the particle arrives at the origin within time $t$

$$\int_0^\infty f_0(t)\,dt' = P(T \le t) = 1 - P(T > t) = 1 - S_0(t) \tag{5}$$

where $T$ represents the random variable corresponding to the first-passage time of the particle. In equation (5) we use the definition of $S_0(t)$ to express the relationship between $S_0(t)$ and $f_0(t)$. Thus from equation (5) one can obtain

$$f_0(t) = -S_0'(t) = \frac{x_0}{\sqrt{4\pi Dt^3}}e^{-x_0^2/4Dt} \tag{6}$$

From equation (6) we computed the values of $f_0(t)$ for a given system and compared them to our numerical results from Monte-Carlo simulation (See Appendix Figure 5).

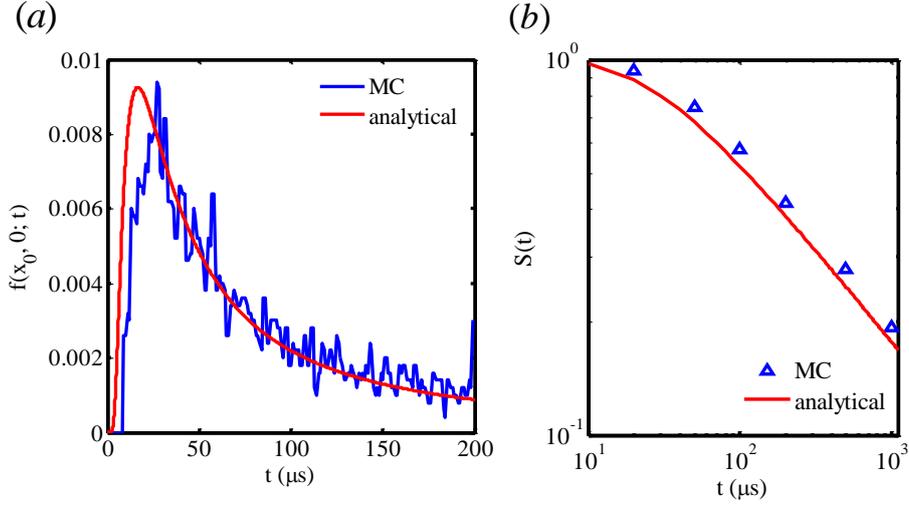

Appendix Figure 5. The comparison between analytical and numerical results of (a) the probability density function of the first-passage time and (b) its survival probability for a given physical system: $D = 10^{-6} cm^2/s$, $x_0 = 0.1 \mu m$.

Now let us extend this problem to a generalized version: What is the distribution of the first-passage time if $N$ particles are simultaneously populated at distinct staring points, $x_1, x_2, \cdots x_N$? First, we define a set of random variables $\{T_i\}$ where $T_i$ indicates a random variable for the first-passage time of the diffusing particle starting at $x_i$ $(i = 1, \cdots, N)$. In addition, we define another random variable $T_{min}$, which is the minimum of the set $\{T_i\}$. Then the survival probability $S^{(N)}(t)$, which is defined as the probability that none of the $N$ particles hits the absorbing boundary by time $t$, can be expressed as

$$\begin{aligned} S^{(N)}(t) &\equiv P(T_{min} > t) \\ &= P((T_1 > t) \bigcap (T_2 > t) \cdots \bigcap (T_N > t)) \\ &= P(T_1 > t) \cdot P(T_2 > t) \cdots \cdot P(T_N > t)) \\ &= S_1(t) S_2(t) \cdots S_N(t) = \prod_{i=1}^{N} S_i(t) \end{aligned} \quad (7)$$

where $S_i(t)$ is the survival probability for an individual particle starting at $x_i$. Consequently one can obtain the analytic expression of the corresponding distribution function of the first-passage time among $N$ particles :

$$f^{(N)}(t) = -\frac{dS^{(N)}(t)}{dt} = -\frac{d}{dt}\left[S_1(t)S_2(t)\cdots S_N(t)\right]$$
$$= -\left[S_1'(t)S_2(t)\cdots S_N(t)\right] - \left[S_1(t)S_2'(t)\cdots S_N(t)\right] - \cdots - \left[S_1(t)S_2(t)\cdots S_N'(t)\right]$$
$$= \left[f_1(t)S_2(t)\cdots S_N(t)\right] + \left[S_1(t)f_2(t)\cdots S_N(t)\right] + \cdots + \left[S_1(t)S_2(t)\cdots f_N(t)\right] \quad (8)$$
$$= \sum_{i=1}^{N} f_i(t)\left[\prod_{j \neq i} S_j(t)\right]$$

Appendix Figure 6 compares the analytical and the numerical solutions of $S^{(N)}(t)$ and $f^{(N)}(t)$ for a given system.

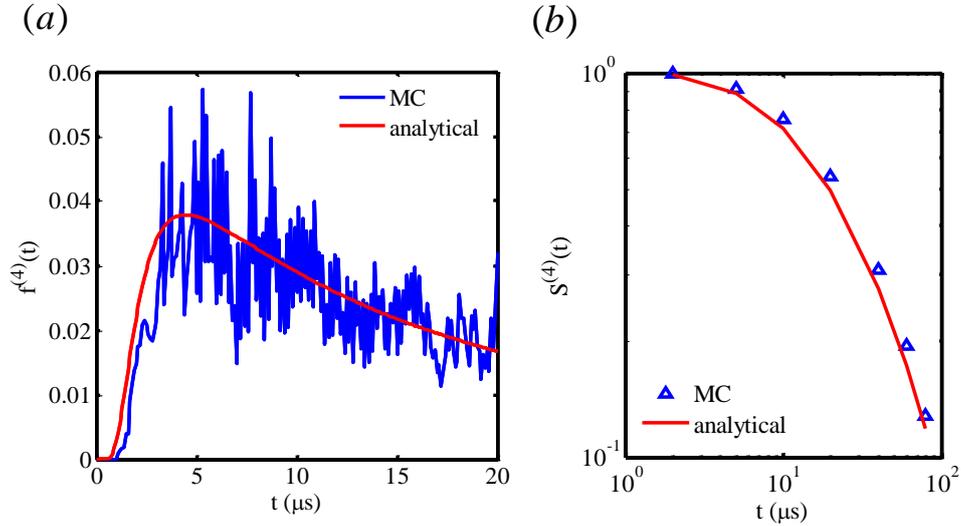

Appendix Figure 6. The comparison between analytical and numerical results of (a) the probability density function of the first-passage time and (b) its survival probability for a given physical system: $D = 10^{-6} cm^2/s$, $x_1 = 0.05 \mu m$, $x_2 = 0.1 \mu m$, $x_3 = 0.15 \mu m$, $x_4 = 0.2 \mu m$. Here totally four particles start their diffusion process at distinct starting points.

II. $k^{th}$-passage time of diffusing $N$ particles in a 1-D semi-infinite system

In the previous section, we developed various analytic expressions for the distribution of the first-passage time. Now we finally want to generalize the problem that can be observed in any practical environment (see Appendix Fig. 1. Of the manuscript): Instead of the first-passage time, how long will the second or third passage time take in $N$ particle system? Can we solve the same problem for the $k^{th}$-passage process?

First, let us consider a simplified case: a one-dimensional system in which all $N$ particles are initially located at $x_0$. If the $k^{th}$-passing particle arrives at the boundary in the time interval $[t, t+dt)$, then this indicates that (1) $k-1$ particles have already arrived at the boundary by time $t$ and (2) there are $N-k$ particles that still remains *alive* by time $t$. Then the formal expression for the probability distribution function for the $k^{th}$ arriving particle to hit the boundary, denoted as $f_{k,N}(t)$, is

$$f_{k,N}(t) = \frac{N!}{(N-k)!(k-1)!}[1-S_0(t)]^{k-1} f_0(t)[S_0(t)]^{N-k}. \tag{9}$$

The factor $[1-S_0(t)]^{k-1}$ is the probability that $k-1$ particles have first-passage times that are less than $t$, while the factor $[S_0(t)]^{N-k}$ gives the complimentary probability that $N-k$ particles have first-passage times that are greater than $t$. The factor $f_0(t)$ is just the probability density function of the first-passage time for the $k^{th}$ time-ordered particle. And finally, the combinatorial factor accounts for the number of distinct arrangements of particle labels that correspond to this $k^{th}$ passage problem. There are $N!$ ways of arranging the particle labels, but we don't care about the labels for all $N-k$ particles that arrive later or all $k-1$ particles that arrive earlier. Thus we have to divide by $(N-k)!$ and by $(k-1)!$. The corresponding survival probability can be simply obtained by integrating $f_{k,N}(t)$ with respect to time. Appendix Figure 7 shows the comparison between the analytical and numerical values of $f_{3,5}(t)$ and $S_{3,5}(t)$.

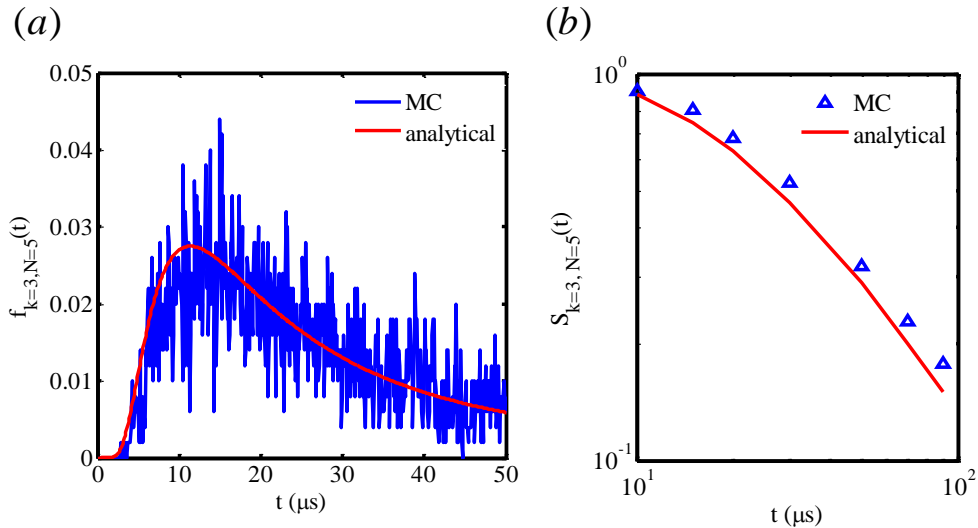

Appendix Figure 7. The comparison between analytical and numerical results of (a) the probability density function of the first-passage time and (b) its survival probability for a given physical system: $D=10^{-6} cm^2 /s$, $x_0 = 0.05 \mu m$. All five particles are initially located at $x_0$.

Next, if we assume that $N$ particles are initially located at distinct positions such that only a single particle is located per starting positions, then just by expanding equation (9) for every initial location, we can obtain the expression for $f_{k,N}(t)$ as

$$f_{k,N}(t) = \sum_{i_1}\sum_{i_2}\cdots\sum_{i_{k-1}}(1-S_{i_1}(t))(1-S_{i_2}(t))\cdots(1-S_{i_{k-1}}(t))\left[\sum_{j\neq i_1,\cdots,i_{k-1}} f_j(t)\left(\prod_{m\neq i_1,\cdots,i_{k-1},j} S_m(t)\right)\right] \quad (10)$$

Finally, let us consider a situation where each of the $N$ particles can be populated at any locations. Of course, there might be more than one particle to be populated at the same location. This is the most-generalized version of the problem we solve in the $k^{th}$-passage process. Even in this case, we can still set up Eq. (10) by simply assuming that all particles are populated at every different location and then modify the equation such that there might be some duplicated terms inside the summations. Appendix Figure 8 shows the comparison between the analytical and numerical values of $f_{3,12}(t)$ and $S_{3,12}(t)$ for a given system in which totally 12 particles are equally distributed to four initial locations.

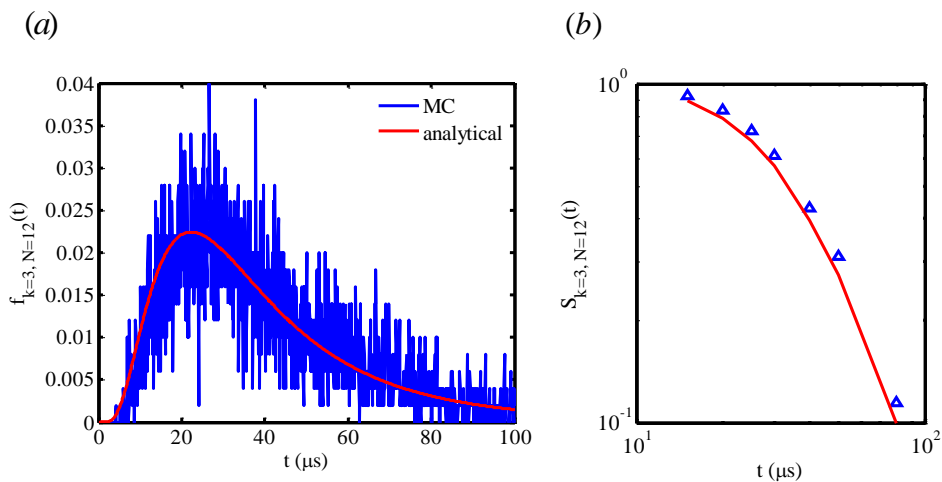

Appendix Figure 8. The comparison between analytical and numerical results of (a) the probability density function of the first-passage time and (b) its survival probability for a given physical system: $D = 10^{-6} cm^2/s$, $x_1 = 0.05\mu m$, $x_2 = 0.1\mu m$, $x_3 = 0.15\mu m$, $x_4 = 0.2\mu m$. Here totally 12 particles are initially located at four distinct positions such that there are 3 particles initially at each position.